\documentclass{LT23auth}
\include{psfig}

\begin{document}

\begin{frontmatter}

\title{Hanbury Brown Twiss effects in channel mixing
normal-superconducting
systems}

\author[address1]{\thanksref{thank1} M. B\"uttiker},
\author[address1]{P. Samuelsson}

\address[address1]{D\'epartement de Physique Th\'eorique, Universit\'e de
Gen\`eve, CH-1211 G\'eneve 4, Switzerland.}


\thanks[thank1]{ E-mail:buttiker@karystos.unige.ch}

\begin{abstract}
An investigation of the role of the proximity
effect in current cross correlations in multiterminal,
channel-mixing, normal-superconducting
systems is presented.
The proposed experiment is an electrical analog of the optical
Hanbury Brown Twiss intensity cross correlation experiment.
A chaotic quantum dot is connected via quantum
point contacts to two normal and one superconducting reservoir. For
dominating coupling of the dot to the superconducting reservoir, a
magnetic flux of the order of a flux quantum in the dot suppresses the
proximity effect and reverses the sign of the cross correlations, from
positive to negative.
In the opposite limit, for a dominating coupling
to the normal reservoirs, the proximity effect is weak and the
cross correlation are positive for a nonideal contact between the dot
and the superconducting reservoir. We show that in this limit the
correlations can be explained with particle counting arguments.
\end{abstract}
%
%
\end{frontmatter}

\section{Introduction}

In normal mesoscopic multiterminal conductors, the cross correlation
between
currents flowing into different terminals are manifestly negative. As
shown in Ref. \cite{Buttiker92}, this is a consequence of the
fermionic statistics of the electrons. Recently, such negative
correlations were observed experimentally \cite{Oliver99}, in an
electrical analog \cite{Buttiker92,Buttiker90,Buttikerrew} of the
Hanbury Brown Twiss experiment \cite{Hanbury56} with photons.

The manifestly negative cross correlations were derived under the
assumption of noninteracting electrons. In contrast,
long range Coulomb interactions between
electrons can give rise to positive cross correlations between
currents at capacitive contacts \cite{Martin00}. Moreover,
in Ref. \cite{Texier00} it is demonstrated
that for inelastic scattering between edge states in a
multiterminal conductor in the quantum Hall regime, the current cross
correlation can be positive. Another situation where positive cross
correlations are possible is when the normal conductor is connected to
a superconducting reservoir \cite{Datta96}.

In a normal-superconducting system, electrons can be converted into
holes via Andreev reflection at the interface between the normal
conductor and the superconductor. The Andreev reflection thus
introduces correlations between electrons and holes in the normal
conductor, a phenomena known as the proximity
effect \cite{Beenakker97,proximity}. The influence of the proximity
effect on the
current auto-correlations, i.e. the shot noise, in two-terminal
normal-superconductor junctions has been studied in Refs.
\cite{Beenakker94,Jehl00}.

In multiterminal conductors, Andreev reflection can lead to positive
cross correlations between currents flowing in the contacts to the
normal reservoirs \cite{Datta96,Torres99}. This was originally
predicted for single mode conductors \cite{Datta96,Torres99}.
In contrast, a discussion of a very asymmetrical geometry in which
one of the contacts is a tunneling tip suggested negative correlations
in the presence of channel-mixing scattering \cite{Grames}.
Interestingly, in Ref. \cite{Nagaev01} it was shown that in a
multiterminal diffusive normal-superconducting system with perfect
interfaces between the normal conductor and superconducting reservoir,
the cross correlations are manifestly negative in the absence of the
proximity effect. Positive correlations can be enforced with the help
of ferromagnetic contacts \cite{Taddei02} a case which we will not
discuss here further.

Clearly, these very different predictions and their sensitivity
to the specific choice of parameters and geometry demand a more
systematic investigation of the role of channel-mixing
and the proximity effect. To investigate the role
of channel mixing
Boerlin et al. \cite{Boerlin02} have analyzed a
geometry with diffusive tunnel junctions. Independently,
the authors \cite{Samuelsson02} present a systematic analysis
of the current correlations in a
system consisting of a chaotic quantum dot connected via point
contacts to one superconducting and two normal reservoirs. Using the
scattering approach for normal-superconducting systems \cite{Datta96}
in combination with a random matrix theory \cite{Beenakker97,Melsen97}
description of the chaotic dot has
the advantage of allowing us to analyze both the case with and without
proximity effect on the same footing. The {\it generic} properties of the
model also makes our results qualitatively relevant for multiterminal
normal-superconducting structures with random scattering.

\section{The model}

\begin{figure}[h]
\centerline{\psfig{figure=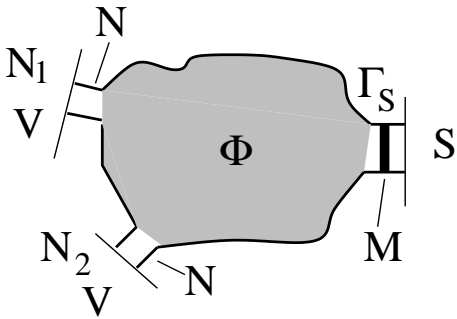,width=6.0cm}}
\caption{A chaotic quantum dot (grey shaded), acting as a
beam-splitter, is connected to two normal reservoirs ($N_1$ and $N_2$)
and one superconducting reservoir ($S$) via quantum point
contacts. There is a magnetic flux $\Phi$ in the dot.}
\label{fig1}
\end{figure}

The three terminal dot-superconductor junction is shown in
Fig. \ref{fig1}. A chaotic quantum dot (see Ref. \cite{Beenakker97}
for definition) is connected to two normal reservoirs ($N_1$ and
$N_2$) and one superconducting reservoir ($S$) via quantum point
contacts. The contact to the normal reservoir is perfect, the contact
to the superconducting reservoir has a mode independent transparency
$\Gamma_S$.

The contacts support $N$ and $M$ transverse modes
respectively. The conductance of the point contacts are much larger
than the conductance quanta $2e^2/h$, i.e. $N,M\Gamma_S \gg 1$, so
Coulomb blockade effects in the dot can be neglected. The two normal
reservoirs are held at the same potential $V$ and the potential of the
superconducting reservoir is zero. The magnetic flux in the dot is
$\Phi$.

\section{Current cross correlations}

Due to the random scattering in the dot, the current $I_i(t)$ in
contact $i$ fluctuates around its quantum statistical average $\bar
{I_i}$. We calculate the zero-frequency spectral density of the current
cross-correlations
\begin{equation}
P_{12}=2\int dt~\overline{\Delta I_1(t)\Delta I_2(0)},
\end{equation}
where $\Delta I_j(t)=I_i(t)-\bar {I_i}$. The correlation $P_{12}$ can
be expressed \cite{Datta96} in terms of the scattering matrix $S$ of
the total system, dot and superconductor. To proceed  $S$ can be expressed
\cite{Beenakker97} in terms of the scattering matrix $S_d$ of the dot
and the Andreev reflection amplitude at the contact-superconductor
interface.

We consider the limit of zero temperature and a potential $eV$ much
lower than the inverse dwell time of the dot, $E_{Th}$, where the
energy dependence of the scattering matrix $S_d$ can be neglected. We
also assume $eV\ll\Delta$, where $\Delta$ is the superconducting
gap. This restricts the transport to an energy interval where no
quasiparticles can escape into the superconductor. In this regime, the
scattering matrix $S$ describes only scattering between the normal
reservoirs.

Noting that the current fluctuation is just the sum of the
fluctuations of electron and hole currents, the noise power can be
conveniently written \cite{Datta96},
\begin{equation}
P_{12}=P^{ee}_{12}+P^{hh}_{12}+P^{eh}_{12}+P^{he}_{12}
\label{totcorr}
\end{equation}
where $P_{12}^{\alpha\beta}$ is the correlation between $\alpha$ and
$\beta$ quasiparticle currents. Unlike the cross correlations in
purely normal systems \cite{Buttiker92}, which are manifestly
negative, the cross-correlation $P_{12}$ can be positive, because the
correlations between different types of quasiparticles,
$P^{eh}_{12}+P^{he}_{12}$, are positive.

\begin{figure}[h]
\centerline{\psfig{figure=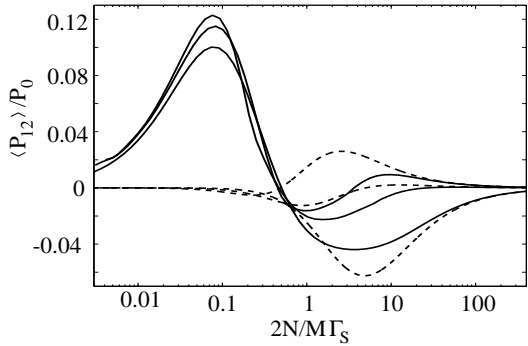,width=7cm}}
\caption{The ensemble averaged current cross-correlation $\langle
P_{12} \rangle$ with (solid) and without (dashed) proximity effect in
the dot, as a function of $2N/M\Gamma_S$. The transparency $\Gamma_S$
is $1.0, 0.8$ and $0.6$ from bottom to top (counting at
$2N/M\Gamma_S=10$).}
\label{fig2}
\end{figure}

The ensemble averaged correlations $\langle P_{12} \rangle$ are
calculated using the statistical properties of the scattering matrix
$S_d$ of the dot \cite{Brouwer95}. The details of these calculations
are presented in Refs. \cite{Samuelsson02,Samuelsson02b}, here we just
show the result in Fig. \ref{fig2} for various transparencies
$\Gamma_S$ of the dot-superconductor contact.

Two different regimes for the magnetic flux are considered. For a flux
much smaller than the flux quanta, $\Phi\ll h/e$, the magnetic flux
has no effect on the proximity effect and can be completely
neglected. In the opposite regime, a magnetic flux larger than the
flux quantum, $\Phi\gg h/e$, effectively breaks the time reversal
symmetry in the dot and suppresses the proximity effect
\cite{Melsen97}.

We see from Fig. \ref{fig2} that suppressing the proximity effect has
a very different effect in the two limits; dominating coupling of the
dot to the normal reservoirs, $M\Gamma_S\ll N$, and to the
superconducting reservoir $M\Gamma_S\gg N$. In the limit of strong
coupling to the superconducting reservoir, $M\Gamma_S\gg N$,
suppressing the proximity effect with a magnetic flux in the dot
completely suppresses the positive correlations.

\begin{figure}[h]
\centerline{\psfig{figure=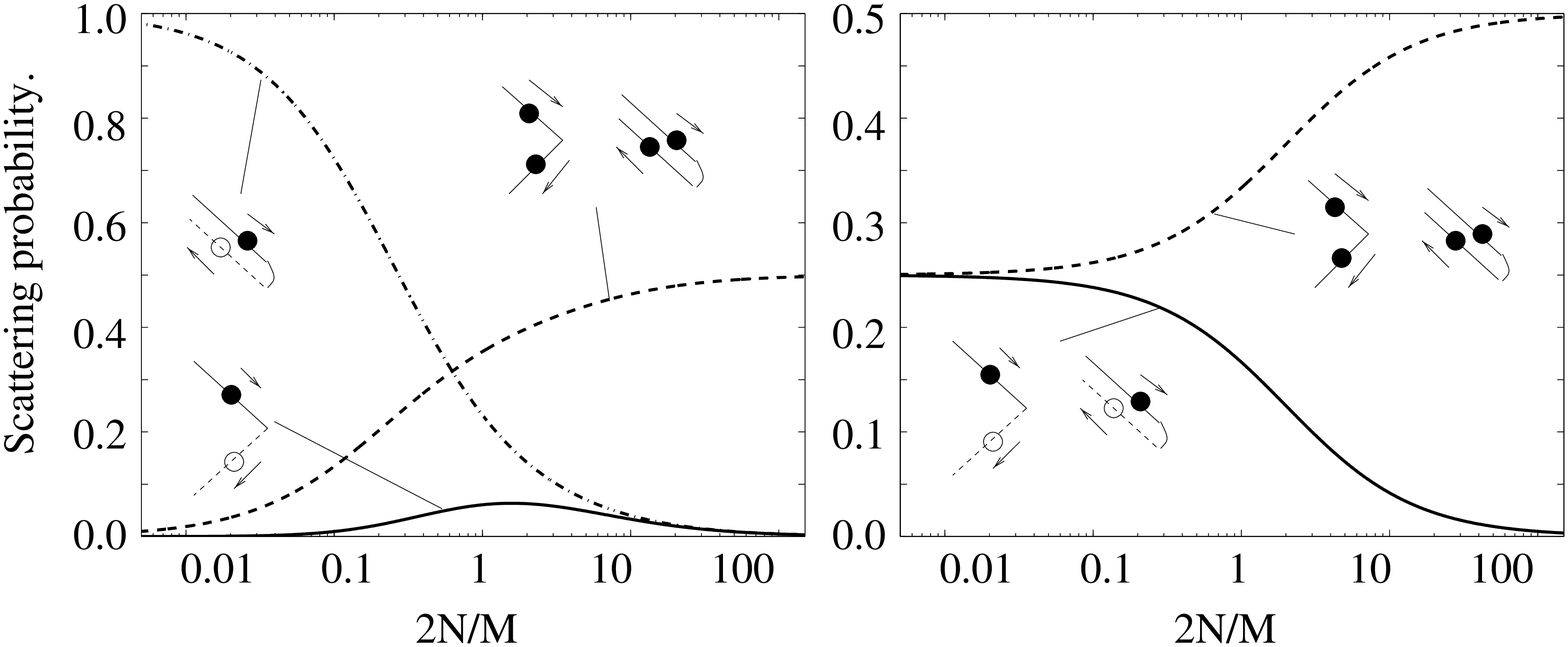,width=7.5cm}}
\caption{The probabilities for different scattering processes for an
electron incoming from reservoir $1$, as a function of $2N/M$ with
(left) and without (right) a proximity effect in the dot. The point
contact to the superconducting reservoir is perfectly transparent,
$\Gamma_S=1$. Filled (empty) circles denote electrons (holes) and
$\nwarrow$ ($\swarrow$) denotes quasiparticles leaving to reservoir
1(2).}
\label{fig3}
\end{figure}

In this limit, in the presence of the proximity effect, there is a gap
in the spectrum around the Fermi energy in the dot \cite{Melsen97}. As a
consequence, quasiparticles injected from one normal reservoir are
Andreev reflected, effectively direct at the contact-dot
interface \cite{Clerk00}, with unity probability back to the same
reservoir. This is shown in Fig. \ref{fig3} for a perfect contact to
the superconducting reservoir, $\Gamma_S=1$, however, the scattering
probabilities (as a function of $2N/M\Gamma_S$) are essentially
independent of $\Gamma_S$. For $2N/M\Gamma_S \rightarrow 0$, the
scattering process is thus deterministic, there is no partition of
incoming quasiparticles and hence no noise, $P_{12}=0$.

Increasing the coupling to the normal reservoir, the probability of
normal reflection as well as cross Andreev reflection, from one
reservoir to the other, becomes finite. As is clear from
Fig. \ref{fig3}, the normal reflection is the dominant process of the
two. On a formal level, we can thus neglect the terms in $P_{12}$
containing the cross Andreev reflection amplitude. As seen in
Fig. \ref{fig3}, this gives $\langle P_{12}\rangle=2 P_0
(N/M\Gamma_S)$, positive since only terms in $\langle
P_{12}^{eh}\rangle+\langle P_{12}^{he}\rangle$ contribute. Here
$P_0=4e^3/h$.

Suppressing the proximity effect, the gap in the spectrum is suppressed
and the scattering processes are strongly modified. This is clearly
seen by comparing the two figures in Fig. \ref{fig3}, showing the
scattering probabilities for $\Gamma_S=1$. In the limit of $2N/M
\rightarrow 0$, an injected electron, which in the presence of the
proximity effect was directly back reflected as hole with unity
probability to the same contact, now has a probability $1/4$ each to
leave the dot in the following four ways: as an electron into contact
1 or 2 or a hole into contact 1 or 2.

As a consequence, the correlations become negative, i.e. there are no
positive correlations in the absence of the proximity effect, similar
to what was found for a metallic diffusive system \cite{Nagaev01}.
Fig. \ref{fig2} demonstrates that this is independent on the transparency
$\Gamma_S$ of the barrier in the dot-superconductor contact.

In the opposite limit, with a weak coupling of the dot to the
superconductor, $M\Gamma_S\ll N\Gamma_N$, the cross correlations are
independent of the proximity effect, as seen in Fig. \ref{fig2}. This
is also observed for the different scattering probabilities in
Fig. \ref{fig3}. An analytical calculation of the correlations in this
limit gives the simple result
\begin{equation}
\frac{\langle P_{12}\rangle}{P_0}=\frac{M}{2N}R_{eh}(1-2R_{eh})
\label{notrssmallar}
\end{equation}
where $R_{eh}=\Gamma_S^2/(2-\Gamma_S)^2$ is the Andreev reflection
probability of quasiparticles incident in the dot-superconductor
contact. There is a crossover from negative to positive correlations,
which occurs for $R_{eh}=1/2$, i.e $\Gamma_S=2(\sqrt{2}-1)\approx
0.83$. In this limit of weak coupling to the superconducting
reservoir, where the proximity effect plays no role, it is possible to
explain the current correlations with straightforward particle
counting arguments.

\section{Particle counting model}

\begin{figure}[h]
\centerline{\psfig{figure=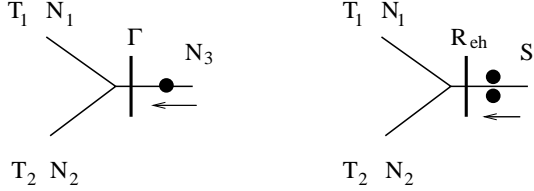,width=7.0cm}}
\caption{Left: Injection of single electrons, corresponding to a
completely normal system. Right: Injection of pairs of electrons,
corresponding to the normal-superconducting system.}
\label{fig4}
\end{figure}

In the limit of weak coupling to the superconductor, a quasiparticle
injected from one of the normal reservoir will at the most scatter
once at the contact between the dot and the superconductor (with a
probability $R_{eh}$ to Andreev reflect), before leaving the dot.
Noting that an Andreev reflection at the dot-superconducting interface
effectively leads to an injection of a pair of particles (electrons or
holes), we can equivalently say that in the limit of weak coupling to
the superconductor, none of the particles in such an injected pair
will ever return to the dot-superconductor contact, but instead
directly and independently leave the dot into one of the normal
reservoirs, with probability $T_1=T_2=1/2$ for each reservoir. This
process is shown schematically in Fig. \ref{fig4}.
Based on this, we can, using particle counting statistics arguments,
calculate the probability $P(N_1,N_2,N_p)$ that $N_1 (N_2)$ electrons
have ended up in the normal reservoir $1(2)$, when $N_p$ pairs have
tried to enter the dot. From the probability $P(N_1,N_2,N_p)$, we can
calculate (the details are presented in Ref. \cite{Samuelsson02b}) the
cross correlations, giving
\begin{eqnarray}
P_{12}=\langle N_1N_2 \rangle-\langle N_1 \rangle\langle N_2
\rangle\propto R_{eh}(1-2R_{eh}).
\end{eqnarray}
This is just the same as was found in Eq. (\ref{notrssmallar}), using
a completely quantum mechanical approach.

As a comparison, considering the injection of single particles (see
Fig. \ref{fig4}), corresponding to the situation when the
superconducting reservoir is replaced by a normal reservoir, we
instead get manifestly negative cross correlations.

From this we can conclude that in the limit of a weak coupling to the
superconductor, the current correlations are determined by the Andreev
reflection at the dot-superconductor interface together with
independent partitioning of particles inside the dot.

\section{Conclusions}
In conclusion, we have investigated the relation between current cross
correlations and the proximity effect in a three terminal chaotic-dot
superconductor junction. We find that both the sign and magnitude of
the correlations can be changed by a weak magnetic field in the dot,
suppressing the proximity effect. In the limit of weak coupling of the
dot to the superconducting reservoir, the current correlation can be
found by particle counting arguments.
%
%
\begin{ack}
We acknowledge discussions with H. Schomerus, Y. Nazarov and
W. Belzig. This work was supported by the Swiss National Science
Foundation
and the program for Materials with Novel Electronic Properties.
\end{ack}

%
%


\begin{thebibliography}{9}
\bibitem{Buttiker92}
M. B\"uttiker, Phys. Rev. B {\bf 46} (1992) 12485.
\bibitem{Oliver99}
M. Henny, S. Oberholzer, C. Strunk, T. Heinzel, K. Ensslin, M. Holland, C. Sch\"nenberger, Science {\bf 284}, (1999) 296, W.D. Oliver, J. Kim, R.C. Liu, Y. Yamamoto, {\it ibid} {\bf 284} (1999) 299; S. Oberholzer, M. Henny, C. Strunk, C. Sch\"onenberger, T. Heinzel, K. Ensslin, M. Holland, Physica E. {\bf 6} (2000) 314.
\bibitem{Buttiker90}
M. B\"uttiker, Phys. Rev. Lett. {\bf 65} (1990) 2901, Th. Martin, R.
Landauer, Phys. Rev. B {\bf 45} (1992) 1742.
\bibitem{Buttikerrew}
For a review, see Ya. M. Blanter, M. B\"uttiker, Phys. Rep. {\bf 336}
(2000) 1.
\bibitem{Hanbury56}
R. Hanbury Brown, R.Q. Twiss, Nature {\bf 177}, (1956) 27.
\bibitem{Martin00}
A. Martin, M. B\"uttiker, Phys. Rev. Lett. {\bf 84} (2000) 3386.
\bibitem{Texier00}
C. Texier, M. B\"uttiker, Phys. Rev. B {\bf 62} (2000) 7454.
\bibitem{Datta96}
M. P. Anantram, S. Datta, Phys. Rev. B {\bf 53} (1996) 16390;
T. Martin , Phys. Lett. A 220, (1996) 137.
\bibitem{Beenakker97}
C. W. J. Beenakker, Rev. Mod. Phys, {\bf 69}, (1997).
\bibitem{proximity}
C.J. Lambert, R. Raimondi, J. Phys. Condens. Matter. {\bf 10} (1998)
901.
\bibitem{Beenakker94}
M.J.M. de Jong, C.W.J. Beenakker, Phys. Rev. B {\bf 53} (1996) 16390;
W. Belzig, Y. Nazarov, Phys. Rev. Lett. {\bf 87} (2001) 067006.
\bibitem{Jehl00}
X. Jehl, M. Sanquer, R. Calemczuk, D. Mailly, Nature (London) {\bf 405} (2000) 50; A.A. Kozhevnikov, R.J. Schoelkopf, D.E. Prober,
Phys. Rev. Lett. {\bf 84} (2000) 3398.
\bibitem{Torres99}
J. Torr\'es, Th. Martin, Eur. Phys. J. B {\bf 12} (1999) 319;
M. Schechter, Y. Imry, Y. Levinson, {\it ibib} {\bf 64}, (2001) 224513.
\bibitem{Grames}
T. Gramespacher, M. B\"uttiker, Phys. Rev. B {\bf 61}, 8125 (2000).
\bibitem{Nagaev01}
K. Nagaev, M. B\"uttiker, Phys. Rev. B {\bf 63} (2001) 081301;
X. Jehl, M. Sanquer, Phys. Rev. B {\bf 63}, (2001) 052511.
\bibitem{Taddei02}
G. B. Lesovik , T. Martin, G. Blatter, Eur. Phys. J. B {\bf 24},
(2001) 287; F. Taddei, R. Fazio, Phys. Rev. B {\bf 65}, (2002) 134522.
\bibitem{Boerlin02}
J. Boerlin, W. Belzig, C. Bruder, Phys. Rev. Lett. {\bf 88}, (2002) 197001.
\bibitem{Samuelsson02}
P. Samuelsson, M. B\"uttiker, Phys. Rev. Lett. {\bf 89}, (2002) 046601.
\bibitem{Melsen97}
J. Melsen, P.W. Brouwer, K.M. Frahm, C.W.J. Beenakker, Physica
Scripta. {\bf T69} (1997) 223.
\bibitem{Brouwer95}
P. Brouwer, Phys. Rev. B {\bf 51} (1995) 16878.
\bibitem{Samuelsson02b}
P. Samuelsson, M. B\"uttiker, cond-mat/0207585.
\bibitem{Clerk00}
A. A. Clerk, P.W. Brouwer, V. Ambegaokar, Phys. Rev. B {\bf 62} (2000)
10226.
\end{thebibliography}
\end{document}